\documentclass[preprint,12pt]{elsarticle}

\pdfoutput=1
\usepackage{epstopdf}
\usepackage{amssymb}
\usepackage{hyperref}

\journal{Journal of Magnetism and Magnetic Materials}

\begin{document}

\begin{frontmatter}



\title{Aharonov-Casher effect and quantum transport in graphene based nano rings: A self-consistent Born approximation}


\author{A. Ghaderzadeh}
\address{Department of Physics, Azarbaijan Shahid Madani University, 53714-161, Tabriz, Iran}
\author{S.H. Ebrahimnazhad Rahbari}
\address{School of Physics, Korea Institute for Advanced Study, Seoul, South Korea}
\author{A. Phirouznia\corref{cor1}}
\ead{Phirouznia@azaruniv.ac.ir}
\address{Department of Physics, Azarbaijan Shahid Madani University, 53714-161, Tabriz, Iran
}
\address{Condensed Matter
	Computational Research Lab. \& Computational Nanomaterials Research Group (CNRG), Azarbaijan Shahid Madani University,
	53714-161, Tabriz, Iran
}

\begin{abstract}
In this study, Rashba coupling
induced Aharonov-Casher effect in a graphene based nano ring is investigated theoretically. The
graphene based nano ring is considered as a central device connected
to semi-infinite graphene nano ribbons. In the presence of the Rashba
spin-orbit interaction, two armchair shaped edge nano ribbons are
considered as semi-infinite leads. The non-equilibrium Green's
function approach is utilized to obtain the quantum transport
characteristics of the system. The relaxation and
dephasing mechanisms within the self-consistent Born
approximation is scrutinized. The Lopez-Sancho method is also applied to obtain the
self-energy of the leads. We unveil that the non-equilibrium current
of the system possesses measurable Aharonov-Casher oscillations with
respect to the Rashba coupling strength. In addition, we have observed the same 
oscillations in dilute impurity regimes in which amplitude of the
oscillations is shown to be suppressed as a result of the relaxations.
\end{abstract}

\begin{keyword}
Aharonov-Casher effect \sep Quantum Transport \sep Graphene \sep nano-ring


\end{keyword}

\end{frontmatter}


\section{Introduction}

Beginning of the twenty-first century has witnessed a tremendous
exposure of graphene both because of its ample applications in science
and technology \cite{castro2010,neto2009,geim2007,geim2009}. This
establishes graphene as an emerging intensive research
field \cite{castro2010}. Noticeably,
low energy electrons in graphene appear as relativistic (Dirac)
particles which behave like massless Dirac fermions\cite{neto2009}. Interestingly, low energy excitations create most of the
intriguing properties of the graphene  \cite{Dietl09}. Many studies aim
at investigation of fundamental properties of
graphene \cite{stankovich,lee2008,saito1992}. Far before its
recognition as an extraordinary material, in a pioneering work Wallac
fromulated the electronic properties of graphene \cite{Wallace47}.  \\

From the structural point of view, graphene can be regarded as a
zero-gap semiconductor or zero-overlap semi-metal\cite{neto2009}. As a one-atom-thick two-dimensional (2D) allotrope of
carbon packed in a honeycomb lattice, graphene is considered as a
basic building block for other graphitic
materials \cite{Geim07,Novoselov04}.  \\

One of the most interesting graphene based structures is the graphene
nano ribbons which can be constructed by cutting strips of graphene
\cite{Ezawa06}. The electronic properties of graphene nano
ribbons depend on the ribbons width, type of the edge, and the
chirality. In Zigzag graphene nano ribbons band structure (ZZ-graphene
nano ribbons), the energy gap becomes zero at the edges of the Brillouin
zone. Furthermore, in armchair graphene nano ribbons (AC-graphene nano
ribbons) the number of dimer lines, $N$, determines the energy band
gap \cite{Wright10}.  \\

Revisiting many standard quantum mechanical effects for graphene has
been a subject of many studies. The Aharonov-Bohm effect is the
modulation of the charge current as a result of the external magnetic
field.  This effect is theoretically revisited for graphene based
quantum rings by Beenakker and Wurm \cite{Beenakker,Wurm}. It is also
experimentally confirmed by Russo et. al, \cite{Russo}.  The
Aharonov-Casher (AC) effect can be considered as dual effect of the
Aharonov-Bohm (AB) effect. The difference between the AB and the AC
effects is that the latter is a consequence of interaction of dipole
moment with an external electric field, however, the former appears as
a result of electric charge interaction with an external magnetic
field.

In this study, the Aharonov-Casher effect has been obtained by the
Rashba spin-orbit interaction (RSOI) which arises as a result of an
electric gate voltage. The RSOI is a controllable interaction that can
be tunably controlled by this gate voltage.  \\

Influence of the spin-orbit interaction in single layer graphene has
been described by Kane and Mele \cite{5,6}. Strength of intrinsic spin-orbit coupling (ISOC) in graphene is very small in comparison with the
RSOI \cite{Rakyta,Gmitra,Huertas,Boettger,Konschuh,Abdelouahed}.
The Rashba coupling strength in graphene, has been predicted to
be up to $0.2eV$ \cite{Dedkov}. This value can be considered as a
relatively high coupling strength for a typical spin-orbit
interaction. Therefore, it can be expected that the Rashba coupling
related effects in graphene will be significantly different from the
similar effects that can arise by this interaction in the other
two-dimensional structures.  \\

In the current study, we have considered a graphene based nano ring as
the central device connected with two semi-infinite armchair leads. We
have studied the Aharonov-Casher (AC) effect caused by the Rashba
spin-Orbit interaction. Influence of the Rashba spin-orbit interaction
can be considered as a process in which the spin of the conduction
electrons are flipped while hopping to the nearest neighboring atoms
\cite{Kane05,Zarea09,Gelderen10}. This interaction results in
oscillation of the electric current as a function of the Rashba
coupling which is known as the Aharonov-Casher
effect \cite{Frustaglia,PhysRevLett}.  \\

We have employed the nonequilibrium Green's function (NEGF) formalism
which is known as the standard approach in the investigation of the
quantum transport in mesoscopic systems and molecular
electronics \cite{Haug08}. The NEGF provides a microscopic theory for
investigation of internal interactions and relaxation
mechanisms \cite{Datta95}. In addition, we have utilized the
Lopez-Sancho approach \cite{Sancho85} for obtaining the the self-energy
of the lead-device coupling. In this way, we are able to treat the
real-space self-energy of the leads. Equipped with the Lopez-Sancho
approach, we are able to calculate surface Green's function instead of
hopping all lead atoms \cite{Dietl09} in a semi-infinite lead within a
fast converging numerical approach \cite{Sancho85}. Meanwhile the
effect of the impurity and lattice vibrations are obtained within the
self-consistent Born approximation in which the self-energy of the
scattering process have been given in the context of the diagrammatic
analysis. Since the lattice vibrations alter the effective hopping
amplitude, we have found that the Rashba interaction is modified in
the presence of the atomic vibrations. Therefore we have considered the effect of the
vibronic couplings at very low Rashba
couplings.  \\

\begin{figure}[t]
  \centering	
  \includegraphics[width=8cm]{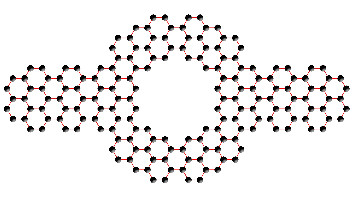}
  \caption{Graphene nano ring connected to two semi-infinite graphene nano ribbons as leads in Armchair shaped edge.}
  \label{fig:01.jpg}
\end{figure}

We have scrutinized dependence of the electric current oscillations to
of the Rashba coupling in pure and impure systems. All above mentioned
calculations have been performed for a graphene based Aharonov-Bohm
system with a central ring shaped structure containing 114 carbon
atoms connected with semi-infinite armchair leads as depicted in
Fig. ~\ref{fig:01.jpg}.
\\


\section{Nonequilibrium Green`s function formalism}
\label{Sec:green_function}



Perturbative formalism, in most of the cases, is the only approach
which could be used to obtain expectation value of an observable in a
many-particle system. Nonequilibrium Green's function approach is an
important technique which can be applied as a perturbative method
especially to obtain the transport properties of open quantum
systems \cite{Datta95}.
The Green's Function of the system is given by
\begin{equation}
G^{(R/A)}(E) = \lim_{\eta \to 0^+} \frac{1}{E\pm i\eta-H},
\end{equation}
where $G^R$ and $G^A$ denote retarded and advanced Green's functions respectively in which $(G^{(R)\dag}(E)=G^{(A)}(E))$, $\eta$ is a small positive value and $H$ is the Hamiltonian of the system.
\\
Therefore the full Hamiltonian can be described as
\begin{equation}
\hat{H} = \hat{H}_{D} + \hat{H}_{L} + \hat{H}_{P},
\end{equation}
where $\hat{H}_{D}$ is Hamiltonian of the central device, $\hat{H}_{L}$ is Hamiltonian of the leads and $\hat{H}_{P}$ denotes the interaction of the electrons with disorders which should be treated perturbatively.
\\
In the current system the effective Hamiltonian of the system in tight-binding approximation is as follow
\begin{equation}
H_D = E_0\sum_ic_i^{\dag}c_i - t\sum_{<i,j>}c_i^{\dag}c_j+\hat{H}_R,
\end{equation}
where $<i,j>$ represents the nearest neighbors and $c_i^{\dag}(c_i)$ is an operator that creates (annihilates) an electron at site $i$ and $t$ is the nearest neighbor hopping in the graphene where we have assumed ($t \approx 2.7 eV$) that fits well with tight-binding approximation. Here it was assumed that the Rashba interaction has been induced by a perpendicular electric field, $E = E \hat{z}$. In this case the Hamiltonian of the Rashba coupling reads \cite{Kane05,Zarea09}
\begin{equation}
\hat{H}_R = it_R \sum_{<i,j>}c^{\dagger}_i \biggl(s\times\mathbf{\hat{d}}_{i,j}\biggr) . \mathbf{\hat{z}}c_j + h.c. .
\end{equation}
Rashba coupling strength, $t_R$, depends on the electric field and increases by increasing this external field, $s$ is the vector of Pauli matrices and $\hat{d}_{ij}$ is the unit vector which connects the $i$ and $j$ lattice sites.
\\
The contribution of the leads is given by the sum of the left and right Hamiltonians denoted by the $\hat{H_L}$ as follows
\begin{equation}
\hat{H_L} = \sum_{i}(\hat{H}^i_L + V^i_{LD} + V^i_{DL}),
\end{equation}
in which $i=$ left or right, labels either side of the ring, $\hat{H}^i_L$ is the Hamiltonian of the $i$-th lead, $V^i_{LD}$ is the hopping between lead and device and similarly $V^i_{DL}$ is the hopping between device and lead where we have $V^i_{DL}=(V^i_{LD})^\dag$\cite{Dietl09}. The Hamiltonian of the leads could be effectively represented by proper self-energies that can be given by
\begin{equation}
\Sigma^R_i = V^i_{DL}g^i_LV^i_{LD},
\end{equation}
where $\Sigma^R_{i}$ stands for the retarded self-energy of $i$-th lead, $g^i_L$ is the Green's function of the isolated semi-infinite lead defined as
\begin{equation}
g^i_L = \frac{1}{E+i\eta-H^i_L}.
\end{equation}
\\
The Lopez-Sancho method \cite{Sancho85} is the approach which provides the self-energy of the semi-infinite leads. Meanwhile it has been shown that the only relevant part of the Green's function of the lead in the transport properties of the system is the surface Green's function which has been considered as a Green's function of the coupling region of the Lead-system\cite{Dietl09}

\section{Aharonov-Casher and Aharonov-Bohm effects}
\label{Sec_AH}
As mentioned before the Aharonov-Casher (AC) effect can be considered
as a counterpart of the Aharonov-Bohm effect. Aharonov-Bohm effect
could be described by the influence of the magnetic field on the
transport of the electric carriers, meanwhile Aharonov-Casher effect
has been considered as an effect in which an external electric filed
has a similar influence on the moving magnetic dipoles e.g.  spin
$1/2$ electrons. In this section we have present a brief description
of the AB and AC effects and their related issues.

\subsection{Aharonov-Bohm effect}

The Aharonov-Bohm (AB) effect \cite{Aharonov59} is a quantum mechanical phenomenon in which an electrically charged particle affected by a magnetic field ($B$), such that the transport of the particle in a field-free region has been effectively controlled by this magnetic filed.
The particle acquires a phase shift while traveling along a given path $P$ as follows
\begin{equation}
\Phi_{AB} = \frac{q}{\hbar}\int_P \mathbf{A}.d\mathbf{r}.
\end{equation}
In which $A$ is the vector potential.
This phase shift results in oscillation of the system current as a function of the magnetic field. Therefore the AB effect can be considered the magneto conductance oscillations in a quantum ring. The AB oscillations\cite{Aharonov59} have been attracted much interests in mesoscopic metallic and semiconductor rings. This effect have been studied  both theoretically \cite{Buttiker83} and experimentally \cite{Levy90} in different conditions. Another important area that AB entered in is the topological effects that is widely studied in recent decade. The AB is a well-known topological effect that has been studied in scattering and bound states \cite{Peshkin89}.

\subsection{Rashba Spin-Orbit Interaction}
Manipulation of electron spin using an external electric field is of crucial importance in nano-scale spintronics \cite{Wolf01} and related technological applications. In the filed of the spintronics both spin and charge of electrons are used to perform data processing. The Rashba spin-orbit coupling is the main candidate of the spin manipulation.
\\
On the other hand due to negligible intrinsic spin-orbit coupling in carbon based systems and long electronic mean paths in graphene \cite{Hongki06,Geim07} graphene becomes a good choice for nearly ballistic spin transport investigations in the new generation of electronic devices.
\\
Kane et. al. showed \cite{Kane05} that the intrinsic SO (ISO) interaction can open a gap in the energy dispersion of mono-layer graphene. Meanwhile the extrinsic Rashba SOI acts in the opposite direction and tends to close the gap. The extrinsic SOI exists only if the lattice inversion symmetry is broken. This symmetry can be broken by the coupling of the graphene sheet with a substrate or by applying an external perpendicular electric field or gate voltage. It has been shown \cite{Gelderen10} that for an external electric field by $E \sim 50V/300$nm, the Rashba coupling is less than $1$ meV\cite{Gelderen10}, impurities can increase the Rashba coupling value up to 7 meV \cite{Castro09} and in one of the studies in this field it was shown that the Rashba coupling can reach up to 0.2 eV \cite{Dedkov}.

\subsection{Aharonov-Casher effect}

In 1984 Aharonov and Casher \cite{Aharonov84} anticipated that when a magnetic dipole has a circular movement in an external electric field particle acquires a phase. This is similar to the Aharonov-Bohm effect in which the phase shift can be acquired by a charged particle in a magnetic field.
In this case the phase shift that was acquired by a particle while traveling along the path identified by $P$ is \cite{Sangster93}
\begin{equation}
\Phi_{AC} = \frac{1}{\hbar c^2}\int_P (\mathbf{E}\times\mu).d\mathbf{r}.
\end{equation}
This effect could be take place even for neutral particles possessing magnetic dipole moment. Therefore these particles exhibit a quantum-mechanical interference effect, which can be considered the dual of the Aharonov-Bohm effect. In both effects the particle acquires a phase shift ($\Phi$) while traveling around external fields. By comparing the AC effect of a neutral particle and the AB effect of a charged particle in the field-free region one can realize that the particle wave packet does not experience an electromagnetic force \cite{Vaidman14} in either of the cases. The AC effect could be considered as a topological effect regarding the relative orientation of the electric field and magnetic moments \cite{Sangster93}. The AB and AC topological quantum effects imply that despite of auxiliary role of electromagnetic potentials in classical electromagnetism they play a relevant physical role in quantum theory \cite{SPAVIERI96}.
\\
The AC phase has been measured experimentally \cite{Sangster93,Cimmino89} with different practical approaches. Measuring the AC oscillations experimentally, exhibits the spin transistor effect which can be used in the silicon nano-sandwiches to indicate the topological basis of the edge channels.
\\
It has been demonstrated that the AC phase shift could be obtained as a result of the spin-orbit interaction (SOI) \cite{Balatsky93}. In general when we apply a perpendicular external field we will have also the Rashba spin-orbit interaction (RSOI) which is a controllable interaction in a system with broken inversion symmetry. Meanwhile the RSOI is a spin-splitting interaction even in zero magnetic field.
\\
In the AB effect the magnetic moment of the charged particle does not precess since the electron is moving in a free-field region. However in the case of the AC effect the spin precesses during the cyclic evolution. This difference led to different transport behaviors induced by the AC phase which reflects itself in spin current induced by the AC phase \cite{Choi97}. There are some anomalous behaviors in the persistent spin currents that come from the change of spin precession angle with varying electric field. There is a persistent spin currents for unpolarized electrons instead of persistent charge currents. The persistent spin current is independent of spin polarization, therefore the required magnetic field for spin polarization is not necessary for the persistent spin current.

\section{The influence of the impurities}

\begin{figure}[t]
  \centering
  \includegraphics[width=7cm]{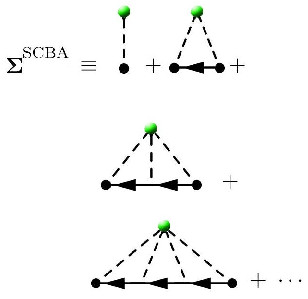}
  \caption{(color online) Diagrammatic expansion of impurity induced self-energy within the self-consistent Born approximation. Thick lines: unperturbed Green's functions; dashed lines: impurity interactions; green points: momentum conserving impurities \cite{Bruus02}.
  }
  \label{fig:SCBA.jpg}
\end{figure}

Ballistic transport falls rapidly in the presence of impurities and electron-impurity interaction breaks the translational symmetry in the mono-layer graphene. Meanwhile impurity induced lattice distortions increases the strength of the SO coupling in graphene. It has been shown \cite{Castro09} that atoms which directly hybridize with a carbon atom induce a distortion of the graphene lattice from $sp^2$ to $sp^3~$. As it was shown in previous investigations of this filed, the electronic density changes as a result of the impurity doping. Accordingly this change of the electronic density, caused by the impurities, leads to a shift in Fermi energy given by $\epsilon_F\approx v_F\sqrt{ n_i}$ in which $n_i$ is the concentration of impurities. Meanwhile impurities lead to an elastic scattering time $\tau_{elas} \approx (v_Fn_i)^{-1}$ and a finite elastic mean free path $l_{elas}\approx an^{-1/2}_i$. Therefore the low impurity regime can be considered as low-density metals \cite{Castro09a} in the dirty limit as $\tau^{-1}_{elas} \approx\epsilon_F$.
\\
Electron-impurity interaction, $V(\mathbf{r},t)$ has been assumed to be a short range potential which build up of randomly distributed and lowly  correlated scattering centers. These conditions could be satisfied by the following impurity profile which satisfies
\begin{eqnarray}
<V(\textbf{x},t)>&=&0\\
<V(\textbf{x},t)V(\textbf{y},t')>&=&\gamma(\textbf{x}-\textbf{y})\delta(t-t')\nonumber\\
\gamma(\textbf{x}-\textbf{y})&=&V^2_{im} e^{-\alpha(\textbf{x}-\textbf{y})^2}.\nonumber
\end{eqnarray}
In which $V_{im}$ is the strength of the impurity potentials.
\begin{figure*}[t]
  \centering
  \includegraphics[width=14cm]{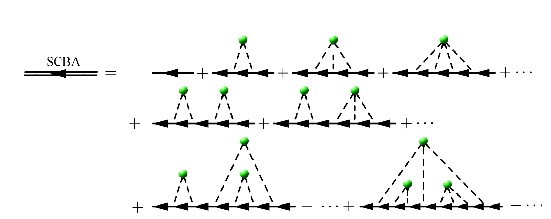}
  \caption{(color online) Averaged Green's function (double line) of the system in the presence of the impurities in term of the free particle (thick lines) Green's function. Dashed lines indicate the impurity interactions}
  \label{Full_SCBA.jpg}
\end{figure*}
Self-energy of the impurities can be given within the self-consistent Born approximation \cite{Haug08}. Diagrammatic expansion of this approach has been presented in Fig.~\ref{fig:SCBA.jpg} in which the averaged Green's function has been given in term of the free particle propagators. Then the self-energy of the impurity interactions is given by
\begin{equation}
\label{self_im}
\Sigma_{im}(x, y, t, t^{'} ) = \gamma(x - y)G(x - y, t, t^{'} )\delta(t - t^{'} ),
\end{equation}
where G is the impurity-averaged Green's function. Therefore this equation provides a self-consistent relation which could be employed to obtain the self-energy of the impurities. Since one can use its counterpart relation which has given by $G[\Sigma]=(E-H_0-\Sigma_{im}[G])^{-1}$.  The diagrammatic representation of the self-energy in the SCBA has been depicted in Fig.~\ref{Full_SCBA.jpg}.
\section{Electron-Vibron Interaction}

\begin{figure}[t]
  \centering	
  \includegraphics[width=9cm]{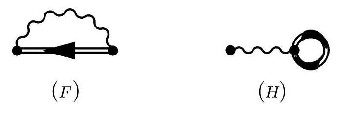}
  \caption{Self-energy of the electron-vibron interaction in the self-consistent Born approximation (SCBA). F stands for the Fock term and H denotes the Hartree term. Double line stands for the averaged Green's function while the wavy lines indicate the electron-vibron interaction.}
  \label{fig:e-v_SCBA.jpg}
\end{figure}

More realistic condition for electronic motion could be achieved when the influence of the atomic vibrations has also been considered in the current formalism for the quantum transport calculations \cite{Mitra08}. Electron-vibron interactions could be described by the following Hamiltonian\cite{Frederiksen04}
\begin{equation}
\hat{H} = \hat{H}_e + \hat{H}_{vib} + \hat{H}_{e-vib},
\end{equation}
where $\hat{H}_e$ describes the electrons, $H_{vib}$ is the Hamiltonian of the free vibron subsystem and $\hat{H}_{e-vib}$ denotes the Hamiltonian of electron-vibron interaction. The atomic vibrations known as vibrons can be described as
\begin{equation}
 \hat{H}_{vib} = \sum_\lambda \hbar \omega_{\lambda} a^{\dagger}_{\lambda} a_{\lambda}.
\end{equation}
In which $a^{\dagger}_{\lambda} $ $(a_{\lambda})$ denotes the creation (annihilation) of vibronic mode $\lambda$ in the system.
At weak electron-vibron interaction strengths the problem can be treated perturbatively, the electron-vibron coupling is described by the following Hamiltonian\cite{Ryndyk06}
\begin{equation}
 \hat{H}_{e-vib} = \sum_{\alpha\beta}\sum_{\lambda} M^{\lambda}_{\alpha\beta}(a_\lambda + a^{\dagger}_\lambda)d^{\dagger}_{\alpha} d_{\beta},
\end{equation}
where $M^{\lambda}_{\alpha,\beta}$ is a hermitian matrix in real-space representation which identifies that the process of electron scattering from a given Wannier-state $\alpha$ to another state $\beta$, which is accompanied by absorbtion or emission of a vibron in $\lambda$ mode. The matrix elements of the electron-vibron interaction have been given by
\begin{eqnarray}
\label{eq5}
\hat{M}^{(\lambda)}=\left(%
\begin{array}{ccccc}
  &0        & m^{(\lambda)} ~~   &  ~~    &   \\
  &   m^{(\lambda)}   ~~& \ddots & m^{(\lambda)} ~~   &   \\
  &     ~~  & m^{(\lambda)}  ~~  & \ddots &  m^{(\lambda)} \\
  &      ~~ &  ~~    & m^{(\lambda)}  ~~  &  0
\end{array}%
\right).
\end{eqnarray}
This matrix indicates that the the effect of the atomic vibrations manifests itself as a reduction in the hopping amplitude.
\\
It was shown that the Rashba spin orbit coupling could be considered as an effective Hamiltonian for the atomic spin-orbit interaction, spin-conserving on-site hopping to the $s$ or $d_{z^2}$ orbitals (via the Stark effect appearing in the presence of an applied transverse electric field) and the hopping originates from the coupling to the $\sigma$-band states \cite{Konschuh, Huertas}. Since the lattice vibrations effectively changes the hopping amplitude accordingly the strengths of the Rashba coupling, could be modified in in the vibrating system. Therefore the quantum transport characteristics of the ring in the presence of the lattice vibrations have been studied at constant and relatively low Rashba couplings. This is due to the fact that the value of the correction of the Rashba coupling as a result of the vibrations increases by increasing the Rashba coupling strengths. This can be inferred from the fact that Rashba coupling linearly increases by the hopping amplitude. Therefore in the presence of the vibronic interactions at low Rashba couplings the value of this correction could be ignored.
\\
Here the atomic vibrations have not been referred as phonon quasi-particles. This is due to the fact that in the absence of the periodic symmetry of the system the Hamiltonian of the vibrations cannot be given by a diagonal matrix in $k$-space representation. In equilibrium state of noninteracting vibrons the vibronic Green's function is
\begin{equation}
D^R_0(\lambda,\omega) = \frac{1}{\omega - \omega_\lambda + i0^+} - \frac{1}{\omega + \omega_\lambda + i0^+},
\end{equation}
\begin{eqnarray}
D^<_0(\lambda,\omega) = -2\pi i \biggl[(N_\lambda + 1)\delta(\omega + \omega_\lambda)+ N_\lambda\delta(\omega - \omega_\lambda)\biggr],
\end{eqnarray}
where $N_\lambda$ is the number of vibrations in the $q$-th mode and it is equivalent to the equilibrium Bose distribution function $f^0_B= 1/(exp(\omega/T) - 1)$ \cite{Ryndyk06}.
Similar to the previous case of the impurity interactions self-consistent Born approximation (SCBA) has been employed for study of the influence of the electron-vibron interaction on the non-equilibrium current transport. In the SCBA the Green's function is determined by the proper self-energy, Fig.~\ref{fig:e-v_SCBA.jpg} within a self-consistent numerical process. The SCBA provides a way to calculate the non-equilibrium distribution functions of electrons and vibrons and bias induced current in the system.
\\
The first Born approximation characterizes by two different lowest order self energy diagrams in the Dyson expansion of the interacting Green's function. These diagrams are known as Hartree and Fock-terms in the Green's function expansion. Self consistent born approximation for this vibronic system could be achieved by replacing the interacting Green's function with bare Green's functions in the Hartree and Fock-diagrams as shown in Fig. ~\ref{fig:e-v_SCBA.jpg}.
\\
In this case self-consistent equations for electron-vibron interaction are given through the relations \cite{Frederiksen04}
\begin{eqnarray}
\Sigma^{H,r}_{SCBA}(\sigma) = i\hbar \sum_{\lambda}\sum_{\sigma^{'}} \frac{2}{\Omega_{\lambda}}\int^{\infty}_{-\infty} \frac{d\omega^{'}}{2\pi}M^{\lambda}\times Tr[G^{<}(\sigma^{'},\omega^{'})M^{\lambda}]
\end{eqnarray}
and
\begin{equation}
\Sigma^{H,\lessgtr}_{SCBA}(\sigma) = 0,
\end{equation}
in which $\Sigma^{H,r}_{SCBA}$ and $\Sigma^{H,\lessgtr}$ are retarded and lesser (greater) self-energies of the Hartree contribution respectively. Similarly the Fock self-energies are given as follows
\begin{eqnarray}
\Sigma^{F,r}_{SCBA}(\sigma,\omega) &=& i\hbar \sum_{\lambda} \int^{\infty}_{-\infty} \frac{d\omega^{'}}{2\pi}M^{\lambda}\times \biggl[D^r_0(\lambda,\omega - \omega^{'})G^{<}(\sigma, \omega^{'})\\
&&+ D^r_0(\lambda,\omega - \omega^{'})G^{r}(\sigma, \omega^{'})
+ D^{<}_0(\lambda,\omega - \omega^{'})G^{r}(\sigma, \omega^{'})\biggr]M^{\lambda}\nonumber,
\end{eqnarray}

\begin{eqnarray}
\Sigma^{F,\lessgtr}_{SCBA}(\sigma,\omega) = i\hbar \sum_{\lambda}\int^{\infty}_{-\infty} \frac{d\omega^{'}}{2\pi}\times M^{\lambda}D^{\lessgtr}_0(\lambda,\omega - \omega^{'})G^{\lessgtr}(\sigma,\omega^{'})M^{\lambda}.
\end{eqnarray}
Meanwhile we have
\begin{eqnarray}
G^{r(a)}_{SCBA}(\sigma,\omega) = \biggl[[G^{r(a)}_{0}(\sigma,\omega)]^{-1} - \Sigma^{r(a)}_{SCBA}(\sigma,\omega) \biggr]^{-1},
\end{eqnarray}
represents the retarded and advanced Green's function in the presence of vibrations and  
\begin{eqnarray}
G^{\lessgtr}_{SCBA}(\sigma,\omega) =
G^{r}_{SCBA}(\sigma,\omega)\Sigma^{\lessgtr}_{SCBA}(\sigma,\omega)G^{a}_{SCBA}(\sigma,\omega).
\end{eqnarray}
It should be noted that
the number of vibrons in the non-equilibrium state could be modified in non-equilibrium regime,
while in an equilibrium state $N_\lambda \equiv N^0_\lambda$. The number of vibrons in a non-equilibrium regime with weak electron-vibron coupling is approximately the same as equilibrium state. The vibron emission, produced by non-equilibrium electrons, changes the non-equilibrium vibronic number $N_\lambda$ which is proportional to the number of electrons \cite{Ryndyk06}.
However in the the self-consistent equations of the current approach, which has been performed via the modification of the Green's functions in a iteration process. The self-consistent process has been considered only on the electronic subsystem.
\section{Current of the Noninteracting Case}

For noninteracting electrons, the use of transmission coefficients within the Landauer-Buttiker formalism to derive the electric current through the system is a well known approach. The Landauer-Buttiker formalism is a well-established method which gives the conductance of the device \cite{Dietl09}. By applying the Fisher-Lee relation \cite{Fisher81}, these coefficients can be written in terms of the system Green's Function. Finally the transmission function between right and left leads is given by
\begin{equation}
T_{i} = Tr[\Gamma_L G_{i} \Gamma_R G_{i}^{\dag}],
\end{equation}
where the function $\Gamma_i$ is the coupling of the device to the leads called the broadening function. Broadening function, $\Gamma_i$, can be obtained in term of the self-energies as follows
\begin{equation}
\Gamma_i= i[\Sigma_i - \Sigma_i^{\dag}].
\end{equation}
Self energy could be considered as an effective interaction which describe the coupling of the device and the leads.
Then the current of the system is given by\cite{VENTRA08}
\begin{eqnarray}
I = \frac{e}{\pi\hbar}\int^{+\infty}_{-\infty}dE\biggl\{[f_l(E) - f_r(E)]\times Tr[\Gamma_rG^+\Gamma_lG^-]\biggr\}.
\end{eqnarray}

\section{Current of the Perturbative Case}

The current formulation derived by Meir and Wingreen \cite{Meir92} relates the electric of current the system to the Green's functions and self-energies in the interacting region. In their formulation the Green's functions and self-energies are obtained in an approximate way.
\\
A noninteracting system coupled to two left and right electrodes can be described by the following self-energies
\begin{equation}
\Sigma^{<}_i (\omega) = in_F(\omega - \mu_i)\Gamma_i(\omega)
\end{equation}
\begin{equation}
\Sigma^{>}_i (\omega) = i\{n_F(\omega - \mu_i) - 1\}\Gamma_i(\omega)
\end{equation}
and
\begin{equation}
G^{\lessgtr}(\omega) = G^R (\omega) \biggl[\sum_i(\Sigma^{\lessgtr}_i(\omega)) + \Sigma^{\lessgtr}_P(\omega)\biggr]G^A(\omega),
\end{equation}
where $\Sigma^{\lessgtr}_P(\omega)=\Sigma_{SCBA}^{(vib)\lessgtr}+\Sigma_{SCBA}^{(im)\lessgtr}$ denotes all of the self-energies sum which effectively give the influence of the scattering mechanisms and perturbations. Where we have defined $G^R=(G^A)^{\dagger}=(E-H_D-\sum_i\Sigma_i-\Sigma_{im}^{r}-\Sigma_{vib}^{r})^{-1}$. In which $\Sigma_{im}^{r}$ and $\Sigma_{im}^{r}$ are the retarded self-energies of disorders and vibrons respectively.
\\

Then the electric current of the $i$-th ($i= Left, Right$) lead is
\begin{equation}
J_i = \frac{1}{\hbar}\int^{\infty}_{-\infty}\frac{d\omega}{2\pi}Tr \bigl[\Sigma^{<}_i(\omega)G^{>}(\omega) - \Sigma^{>}_i(\omega)G^{<}(\omega)\bigr].
\end{equation}

It should be noted that in the present study it was assumed that there is no correlation between the impurity and vibronic relaxations. This means that the impurity relaxation potential could not be modified by lattice vibrations. Therefore to avoid any artificial correlation between the impurity interactions and vibronic couplings the self-energy of these relaxations have been obtained in different and independent self-consistent loops.

\section{Results and Discussion}
\begin{figure}[t]
\centering
\includegraphics[width=8.25cm]{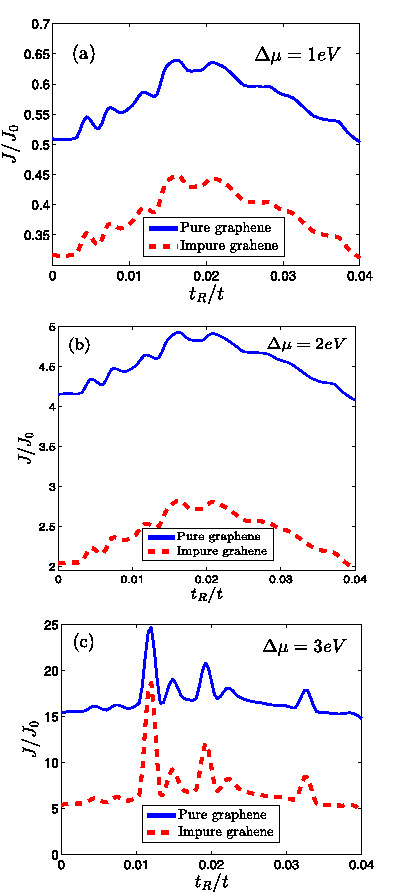}
  \caption{The conductance oscillations for pure and impure vibrationless ($m^\lambda=0$) armchair system at different bias voltages ($\Delta\mu $). The Rashba coupling strength varies in the range: $t_R = 0.0-0.04 t eV$.}
  \label{Impure}
\end{figure}

\begin{figure}[t]
  \centering	
  \includegraphics[width=9cm]{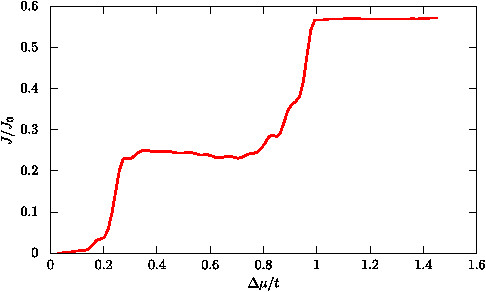}
  \caption{I-V curve of the graphene ring in the presence of the impurity and lattice vibrations at $t_R = 0.013 eV$.}
  \label{ev1}
\end{figure}

\begin{figure}[!h]
  \centering	
  \includegraphics[width=9cm]{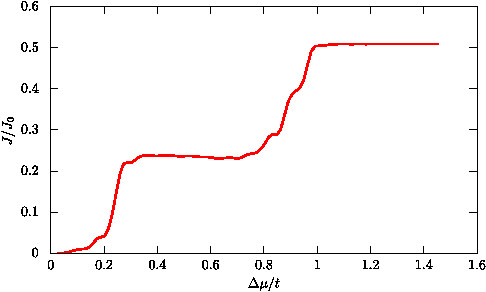}
  \caption{I-V curve of the graphene ring in the presence of the impurity and lattice vibrations at $t_R = 0.0 eV$}
  \label{ev2}
\end{figure}

By means of numerical analysis, we have computed the non-equilibrium
current in graphene based quantum rings at different interaction
regimes. The results of the current study show several cases of conductance oscillation for
pure graphene. We have also obtained the transport features in the presence of the weak dephasing
mechanisms of lowly correlated impurities.  In this study
impurity correlation function adopted with $\alpha = 10 a^{-2}$ and
$V_{im} = 0.01 t$ in which $a$ is the carbon-carbon distance. To
specify the considered electron-vibron interaction we have assumed a
single mode vibrating system ($\lambda=0$) in which the frequency of
the atomic vibrations is $\omega_0 = 0.1 eV$. In addition we have
assumed that $m =
\frac{\omega_0}{100}$. $e\Delta\mu=e\mu_{L}-e\mu_{R}$ is the
difference of the electrochemical potential of the left and right
leads which equals the applied bias voltage. The first nearest
neighbor hopping amplitude for graphene ($t \approx 2.7 eV$) is
considered the energy unit in numeric calculations. Then the current
of the quantum ring have been obtained for different bias voltages and
Rashba couplings. In what follows current of the system has been given
in term of the current unit which has been selected $J_0=et/\hbar$.
\\ In this study the effects of the impurity and electron-vibron
interactions as two perturbative couplings have been considered within
the SCBA in which the effect of these interactions have been given by
two different self-energies. By above specifications current-voltage
curves have been calculated for pure and disturbed systems. Results
show the conductance oscillations in both of the cases.
\\
The conductance oscillations in pure system, i.e. in the absence of the perturbations, are calculated for the graphene based quantum ring connected with the armchair leads as a function of the Rashba coupling strength (Fig. ~\ref{Impure}).
\\
Current oscillation as a function of the Rashba coupling strength could be explained by the phase change acquired by an electron in gate induced Rashba filed within the adiabatic regime. It should be noted that the condition of the occurrence of the geometrical phase shift has been generalized to non-adiabatic  transport. Aharonov and Anandan demonstrated that the geometric phases are not limited to adiabatic regime \cite{aa}. As suggested in some of the previous works of this field, the spin precession rate can precisely be manipulated by a gate voltage generating the Rashba interaction. Datta-Das spin FET has a similar functionality in which ferromagnetic source and drain leads should be employed \cite{nitta}.
\\
In the presence of the impurities bias induced electric current has also been obtained as a function of the Rashba coupling strengths (Fig. \ref{Impure}). The effect of the impurities has been obtained at different bias voltages as depicted in Fig \ref{Impure}. As shown in this figures impurities suppresses the amplitude of Aharonov-Casher oscillations in the quantum ring. It can be inferred from the fact that in the current system impurities act as both relaxing and dephasing factor in the transport process. Meanwhile as shown in this figure, impurity induced dephasing at the limit of weak impurity potentials is less effective since the AC oscillations are measurable in both of the cases. The functionality of the AC oscillations has been preserved in this limit. However, as it was shown in the figures the overall electric current in the system decreases in the presence of the impurities. Results also show that in both of the pure and impure graphene rings the amplitude of the AC oscillations drastically increases at high bias voltages.
\\
Finally it can be inferred that the conductance ($dI/dV$) Aharonov-Casher oscillations due to the Rashba spin-orbit interaction gives pronounced AC in the case of electron gas system compared to the present case of honeycomb lattice \cite{hatano2007non,chen2011tunable,xing2010controlled}. However it should be considered that for a meaningful comparison the number and nature of transverse modes plays an important role in transport characteristics of the different systems.     
\\
It should be noted that in the current study we have ignored the effect of the impurities on the strengths of the Rashba interaction. Since the hopping amplitude is effectively controlled by the system over all local field which build up by effective contribution of the impurities. Therefore this would suggest that the Rashba coupling the strengths could be modified by the impurities. In the current work we have ignored this possibility and the strengths of the Rashba coupling has been assumed to be independent of the impurities.
Finally we have obtained the I-V curve of the quantum ring in the presence of the relaxations at low Rashba couplings as shown in Figs. \ref{ev1} and \ref{ev2}. It was realized that the vibronic effect is much more stronger than the influence of the impurities at least within the self-consistent Born approximation and current transport regime. In this case we have obtained the current of the system when the effects of the atomic oscillations on the hopping amplitude could be considered low enough where the modification of the Rashba interaction due to the lattice vibrations could be ignored. As it can be inferred from these figures that the Rashba coupling increases the non-equilibrium current in the quantum ring at relatively high bias voltages. 
\\
\\
The effect of the Rashba interaction could be considered from two different points of view. First as mentioned before the Rashba coupling could introduce a phase shift in the coherent regime or even when the dephasing mechanisms are relatively weak. In this case increasing the Rashba coupling could increase the electric current. On the other hand the effect of the Rashba interaction could be considered when the system has been moved by the relaxations and the dephasing mechanisms far from the coherent regime. In this case the Rashba coupling induced phase shift and spin switching effects will be suppressed. It should be noted that the Hamiltonian of the Rashba interaction commutes with momentum operator and therefore Rashba interaction itself can not be considered as a scattering mechanism. Meanwhile due to the splitting effect which arises as a result of the presence of this spin-orbit coupling, the spin resolved population could be effectively change near the Dirac points. The change of the population of the current carrying states could decrease the current of the system. This was in apparent contradiction with first case in which the increased electric current Rashba coupling induced phase shift.
\\
Therefore the result of the current study which indicates the increment of the electric current by increasing the Rashba coupling suggests that in the limit of the weak atomic vibrations system could still be considered coherent and the increased current should be referred to the Rashba coupling induced phase shift.
\\
However it should be noted that reduction of the current as a result of the increasing the Rashba coupling could also take place in the coherent regime when the system current has been modulated by the Rashba coupling strength. The main difference between these two regimes is due the fact that the occupation number of the spin bands which are located near the Fermi energy monotonically decreases by the Rashba coupling and this results in monotonic reduction of the system current. Meanwhile in the coherent regime, current oscillates as a function of the Rashba coupling due to the phase shift introduced by this coupling.
\\
Therefore the distinction between these two cases could be appeared when we can obtain the Rashba coupling strengths as a function of the amplitude of the atomic vibrations. Since as it was previously discussed in this work the Rashba coupling strength depends on the hopping amplitude which could be effectively modified by the vibrational motions. In this case we can obtain the system current as a function of the Rashba coupling in a vibrating system. Then we can decide that how far the coherent regime holds by increasing the electron-vibron interaction strength. Since the coherent regime manifest itself in the current oscillations as in term of the Rashba coupling.



\section{Acknowledgment}
This research has been supported by Azarbaijan Shahid Madani university.\\
\\
 \bibliographystyle{elsarticle-num}
 \bibliography{Paper}




\end{document}